\documentclass[]{interact}

\usepackage{epstopdf}
\usepackage{subfigure}

\usepackage{natbib}
\bibpunct[, ]{(}{)}{;}{a}{}{,}

\theoremstyle{plain}

\theoremstyle{definition}

\theoremstyle{remark}

\usepackage{color}
\usepackage{fancyvrb}

\DefineVerbatimEnvironment{Highlighting}{Verbatim}{commandchars=\\\{\}}
\usepackage{framed}
\definecolor{shadecolor}{RGB}{248,248,248}

\usepackage{hyperref}
\usepackage[utf8]{inputenc}

\usepackage{amsfonts}
\usepackage{graphicx,xfrac}
\usepackage[font=normal,labelfont=bf]{caption}
\usepackage{amsgen,amsmath,amstext,amsbsy,amsopn,amssymb,bbm}

\begin{document}

\title{The Hall of Fame Cut in Major League Baseball}

\author{\name{Shane T. Jensen$^{a}$}
\affil{$^{a}$Department of Statistics and Data Science, The Wharton School, University of Pennsylvania, Philadelphia, PA, USA; stjensen@wharton.upenn.edu}}


\maketitle

\begin{abstract}
I present a simple and transparent standard for career greatness in baseball: any major league player with H $> 2500$ or HR $> 350$ or K $> 2800$ or W $> 240$ makes my Hall of Fame Cut.  Rate statistics are avoided due to small sample issues and to ensure the standard is permanent once achieved. Hits and home runs were chosen to represent the two extremes of batting styles.  Strikeouts are chosen as the most fundamental unit of pitching performance whereas wins are included in deference to their historical importance as a benchmark.    Most major league batters and pitchers in the elected Hall of Fame also make my Hall of Fame Cut but my quantitative standard shifts attention to several under-appreciated players, such as Johnny Damon and Bartolo Colon, and allows us to celebrate recent and active players without the waiting period (5 years post-retirement) needed for Hall of Fame election.   My Hall of Fame Cut is also agnostic to performance enhancement or off-field issues and strongly favors longevity over peak performance. 
\end{abstract}

\section{Introduction}

In this paper, I create a new quantitative standard for the baseball Hall of Fame that is unambiguous while still having enough nuance to capture different types of baseball performance.  My goal is a empirical Hall of Fame standard that is transparent, not model-based, and easy to describe.   

Note that this goal is very different than the goal of predicting which major league players will be elected to the Hall of Fame using the best empirical methods and data available.  For this more complex task, I suggest the Jaffe WAR Score system \citep{jaws} which is based on accumulated Wins Above Replacement \citep{war}.  Wins Above Replacement is the industry standard as a single measure of baseball performance but it is not a transparent or simple calculation and thus not well suited for my goal.  Alternative approaches for this more complex task include artificial neural networks \citep{YouHolWec08} and random forests \citep{Fre10}.

In contrast, I am seeking a simple enough decision rule that it can be recalled and applied with ease. I refer to this goal as ``bar napkin simple", i.e. simple enough that the entire selection algorithm can be written on a single bar napkin.  I was inspired in my approach by the Meikyukai\footnote{\url{https://meikyukai.jp/}}, or Golden Players Club, for Japanese baseball where players are automatically inducted upon achieving a certain number of career hits or career wins (saves was added later as a third total).  

I take a similar approach to setting my quantitative standard for career baseball greatness which I will refer to as the {\it Hall of Fame Cut}.  The term ``cut" references the fact that my Hall of Fame standard is defined by thresholds on measures of career hitting or pitching performance.  The need to incorporate multiple aspects of batting and pitching performance while still keeping our selection process ``bar napkin simple" motivates our HOF cut to based on two career counts each for batters (hits and home runs) and pitchers (strikeouts and wins).  

\begin{figure}[ht!]
\includegraphics[width=4in]{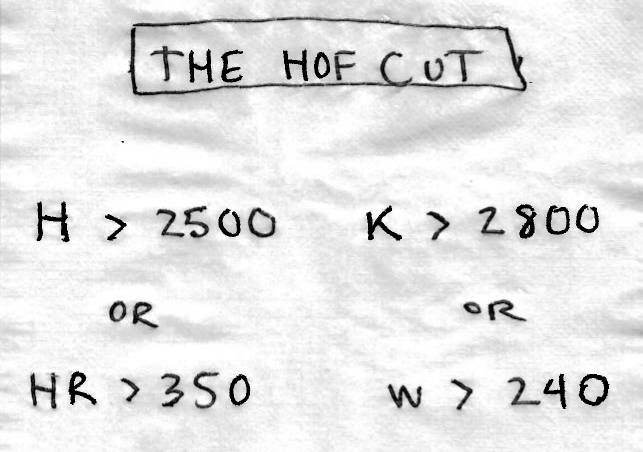}
\centering
\caption{Pictorial representation of the Hall of Fame Cut, hand drawn on a bar napkin.}
\label{fig-cut}    
\end{figure}

Specifically, a batter makes the {\it Hall of Fame Cut} if they have over 2500 career hits or 350 career home runs.  A pitcher makes the {\it Hall of Fame Cut} if they have over 2800 strikeouts or 240 wins.   These thresholds were chosen to give roughly the same number of players in my HOF cut as the current elected HOF but in Section~\ref{landscape} we explore the ``landscape" of different thresholds on career counts.  In Figure~\ref{fig-cut}, I give a pictorial of my Hall of Fame Cut which illustrates the "bar napkin" simplicity of my standard for career greatness in baseball.  

In Section~\ref{whatstats}, I give my rationale for choosing hits, home runs, strikeouts and wins as the particular statistics that I use to represent baseball performance. Then I explore the landscape of possible thresholds for each of those statistics and then choose specific thresholds to roughly match the size of the current elected Hall of Fame in Section~\ref{landscape}.  I compare the set of players that make my Hall of Fame Cut to the set of players in the elected Hall of Fame in Sections~\ref{comparing-batters} and~\ref{comparing-pitchers} and then I create even more selective thresholds to define a "top tier" within the Hall of Fame in Section~\ref{toptier}.  I examine recent and active batters and pitchers in Sections~\ref{recent-and-active-batters} and~\ref{recent-and-active-pitchers} and then conclude with a brief discussion in Section~\ref{discussion}.  

\section{Data and Issues with Elected Hall of Fame}

The Lahman Baseball Database \citep{lahmandatabase} was used for all quantitative analyses that lead to my Hall of Fame Cut.  These analyses were done in R where the Lahman database is available directly as a library in the R software package \citep{R21}.  When my analyses were performed in Fall 2024, the Lahman R library contained player data up to the end of the 2022 season.  Elected Hall of Fame information is also contained in the Lahman database though manual curation was needed to match the 273 players listed on the Baseball Hall of Fame website\footnote{\url{https://baseballhall.org/}}.  For example, the fielding position for each player needed substantial manual curation since fielding information is limited in the Lahman database.  

A larger issue with any quantitative analysis of the Hall of Fame is the lack of data for Negro league players.  So for the purposes of my quantitative analyses, I will assume the 28 elected Negro League players remain in the Hall of Fame automatically, and we will focus our analysis on a quantitative Hall of Fame standard specifically for the major leagues.  There are 245 major league players in the elected Hall of Fame: 74 pitchers and 171 batters.  My Hall of Fame Cut given in Figure~\ref{fig-cut} is designed to have a similar number of major league pitchers and batters above the cut.   

As my goal is a standard for career greatness, I will focus only on cumulative quantitative measures of baseball performance, i.e. counts.  For every major league batter that played from 1871-2022, we collected their career count of games played (G), at bats (AB), runs (R), hits (H), home runs (HR) and runs batted in (RBI).  For every major league pitcher over that same time span, I collected their career count of games played (G), wins (W), strikeouts (K), and total outs.

\section{What Baseball Statistics to Use?} \label{whatstats}

I will consider only the counting statistics listed above as measures of career performance.  Most predictive analyses of either hitting or pitching in baseball focus on rate statistics: BA, OBP, ERA, WHIP, etc.  However, there are at least two reasons why rate statistics are not useful for establishing a simple career HOF standard.  

The main problem is the small sample issue: some players have the most extreme rate values only due to small numbers of opportunities.  For example, there are a myriad of major league players in baseball history with perfect career 0.00 ERAs and less than 10 innings pitched.  So I would need to set additional thresholds on the number of opportunities for each rate statistic used, and Figure~\ref{fig-cut} would need to include those additional choices (e.g. min 1000 PA).  Count statistics by their very nature do not have these small sample difficulties.

An additional problem with using a rate statistic for the Hall of Fame Cut is that Hall of Fame inclusion should be permanent once achieved.  A hitter that surpasses a particular threshold on a counting statistic (e.g. career hits more than 2500) will stay above that mark for all time.  However, a hitter that surpasses a particular threshold on a rate statistic (e.g. career OBP more than 0.4) could then continue to move above and below that mark for the rest of their career which seems unsatisfactory as a Hall of Fame standard.  In contrast, a Hall of Fame cut based upon counting statistics such as in Figure~\ref{fig-cut} has the satisfying property that players can be celebrated for achieving the standard (permanently) while their playing career is still ongoing. 

Having established that my focus will be on the counting statistics listed above, the first empirical question that I will explore is: ``which career counts are the most associated with inclusion in the elected Hall of Fame?"    

As noted above, there are 245 major league players in the elected Hall of Fame: 74 pitchers and 171 batters.  If we were to form a new equally sized Hall of Fame based on just a single batting statistic (e.g. the top 171 batters by career hits), we can tabulate the number of elected HOFers in this top 171 as the ``selectivity" of a particular batting count for the elected Hall of Fame.   

The top 171 batters ranked by hits contains 105 elected HOFers, compared to 108 elected HOFs for batters ranked by runs, 95 elected HOFs for batters ranked by runs batted in, and 60 elected HOFers for batters ranked by home runs.   We can interpret these results as hits and runs being slightly more selective (as single measures) for elected HOFers than runs batted in and substantially more selective than home runs.  

Similarly, we can tabulate the number of elected HOFers in the top 74 pitchers ranked by a pitching count as the ``selectivity" of that pitching count for the elected Hall of Fame.  The top 74 pitchers ranked by wins contains 48 elected HOFers, compared to 36 elected HOFs for pitchers ranked by strikeouts, and 44 elected HOFers for pitchers ranked by total outs.  It is interesting that wins are the most selective of the elected Hall of Fame when strikeouts and total outs are more directly under the control of the pitcher.  

But even the best single measure can not capture the multi-faceted nature of baseball greatness.  And so I define my Hall of Fame standard based on two measures, career hits and career home runs, to capture batting performance and two measures, career strikeouts and career wins, to capture pitching performance.

My rationale for choosing hits and home runs is that these two measures together represent the two extremes of the power spectrum of batting outcomes, which gives a path to the Hall of Fame for different hitting styles (e.g. Ichiro Suzuki versus David Ortiz).  Runs and RBIs also have value as a measure of a player's contribution but much more impacted by teammate performance compared to hits and home runs.  

Also my examination of single measures above suggests that there is not as much redundancy between hits (105 elected HOF based just on hits) and home runs (60 elected HOF based just on home runs) than there would be if the standard was based on runs and RBIs (108 vs. 95 elected HOF based on just runs vs just RBIs).  
There is only so much that one can capture with just two measures but one could consider including walks or stolen bases as part of a more complex HOF standard.  

My rationale for choosing career strikeouts for pitchers is that the strikeout is the most fundamental unit of isolated pitching success (similar to the hit for batters).   Though catchers still play a role in the strikeout, the strikeout is still more isolated to the pitcher than total outs which are also impacted by teammate fielding.  

For a second pitching measure, I choose career wins mostly as a nod to the historical importance of wins as the standard for pitching greatness.  It's importance is supported by the finding above that career wins is the single pitching measure that is most selective for the elected Hall of Fame.  However, wins are controversial as a measure of pitching ability because of the substantial impact of teammate and opponent performance, especially compared to the simple and isolated contribution of strikeouts.  

Quality starts are an attempt to more directly measure pitcher contribution on the game level.  However, quality starts are only tabulated for recent years and thus can not be used as a Hall of Fame standard.  

Any other counts that have been tabulated for all major league baseball players could be considered for a Hall of Fame Cut, but now that I have decided on career hits, home runs, strikeouts and wins as my summaries of batting and pitching performance, I will move onto my rationale for the particular thresholds given in Figure~\ref{fig-cut}.  

\section{Joint Landscapes of Possible Thresholds}\label{landscape}

Now that I've chosen career hits, home runs, strike outs and wins as my measures of batting and pitching greatness, the other task in specifying my Hall of Fame Cut is choosing the thresholds for each of these counts, i.e. the "cut" line for the Hall of Fame.  Ultimately, the decision about what set of thresholds to use is equivalent to the decision of how many players you want in the Hall of Fame, and so we will think about current Hall of Fame size as the outcome of this task.
  
The thresholded nature of this framework suggest that tree partition approaches could be useful.  However, it should be noted that the OR structure of our Hall of Fame Cut decision (e.g. 2500 hits or 350 home runs) does not fit the tree framework where nested variables represent AND relationships.  

Instead, our strategy will be to explore the {\it landscape} of HOF sizes for a set of all reasonable thresholds on career hits, homeruns, strikeouts and wins. I will first examine the Hall of Fame landscape for batters jointly over different thresholds on hits and homeruns and then examine the Hall of Fame landscape for pitchers jointly over different thresholds on strikeouts and wins.   

These joint landscapes will then be used to pick specific thresholds for each of the 4 measures that give a HOF size that roughly matches the size of the elected Hall of Fame (74 pitchers and 171 batters).  I will also explore the sensitivity of the Hall of Fame cut to small increases or decreases in the thresholds and mention some specific players on the margins. 

\begin{figure}[ht!]
\includegraphics[width=5.5in]{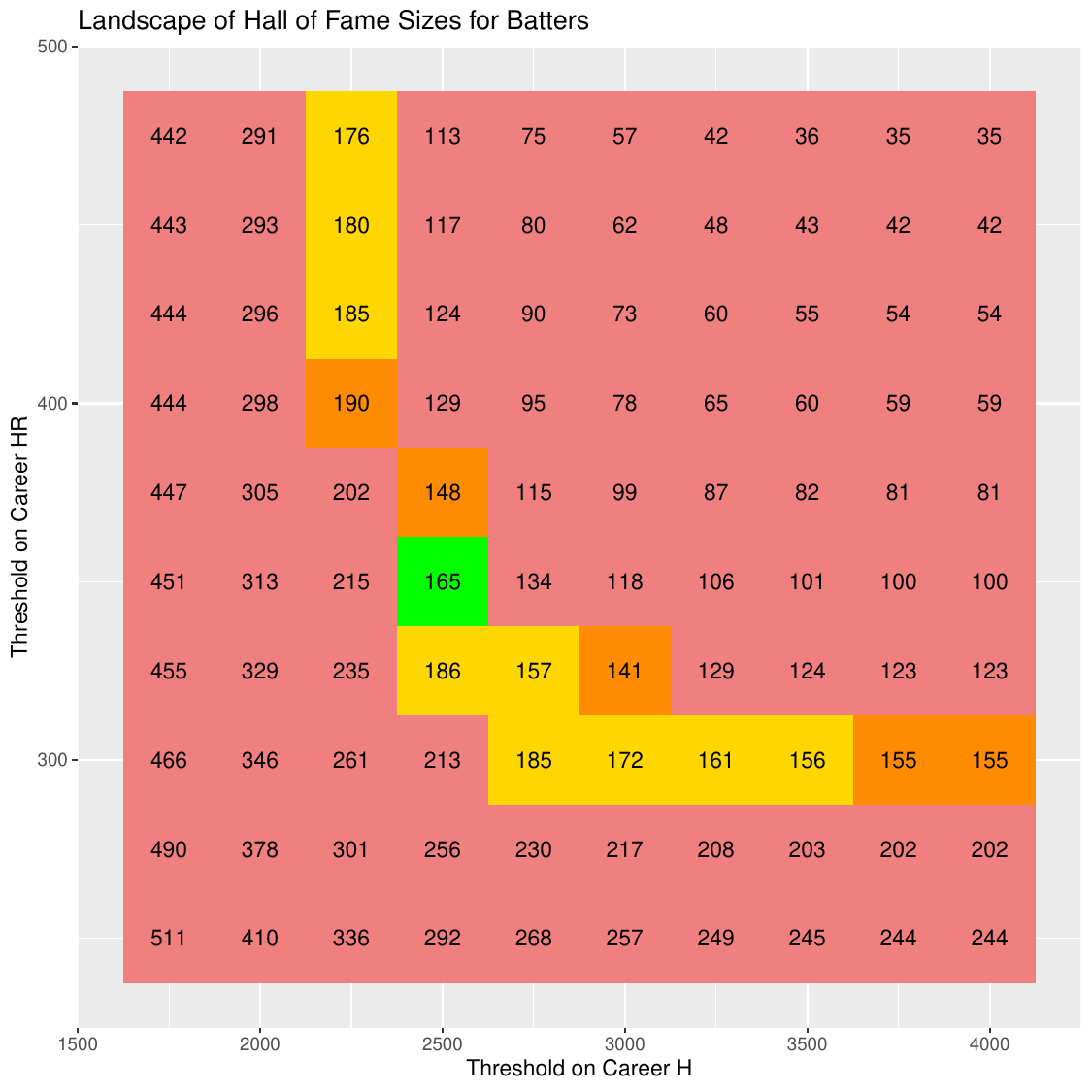}
\centering
\caption{Joint landscape of Hall of Fame sizes for combinations of different thresholds on hits and home runs.  Yellow combinations are near to the size of the elected HOF (171 batters).}
\label{fig-landscape-batters}    
\end{figure}

Figure~\ref{fig-landscape-batters} gives the joint landscape of HOF sizes for a grid of reasonable thresholds on career hits and home runs for batters.  The set of possible thresholds was chosen to span a wide range of elite career values while only considering round numbers so that the resulting Hall of Fame cut is easy to remember. 

We see that several different combinations of thresholds on career hits and home runs lead to Hall of Fame sizes that are close to the 171 batters in the elected Hall of Fame.  Within those options, I chose the thresholds of 2500 hits and 350 HRs that are given in Figure~\ref{fig-cut} which lead to a Hall of Fame wuth 165 major league batters which will be examined in more detail in Section~\ref{comparing-batters}. 

Compared to other combinations in Figure~\ref{fig-landscape-batters} that give similar HOF sizes (e.g. 3000 hits and 300 HRs), my selected thresholds of 2500 hits and 350 HRs gives more balance between hits and home runs in that roughly the same number of batters exceed the hit threshold versus exceed the home run threshold.  

If we were to raise the threshold on career hits from 2500 to 2750, we would reduce the HOF size from 165 batters down to 134 with notables such as Joe Morgan and Roberto Alomar among the 31 additional batters that would be excluded.  If we were to raise the threshold on career home runs from 350 to 375, we would reduce the HOF size from 165 batters down to 148 with notables such as Joe DiMaggio and Yogi Berra among the 17 additional batters that would be excluded. 

If we were to lower the threshold on career hits from 2500 to 2250, we would substantially increase the HOF size from 165 up to 215, with some notables such as Ron Santo, Ozzie Smith and Kirby Puckett among the 50 additional batters that would be included.  If we were to lower the threshold on career home runs from 350 to 325, we would increase the HOF size from 165 up to 186, with Hank Greenberg and Ron Santo (again!) among the 21 additional batters that would be included.

\begin{figure}[ht!]
\includegraphics[width=5.5in]{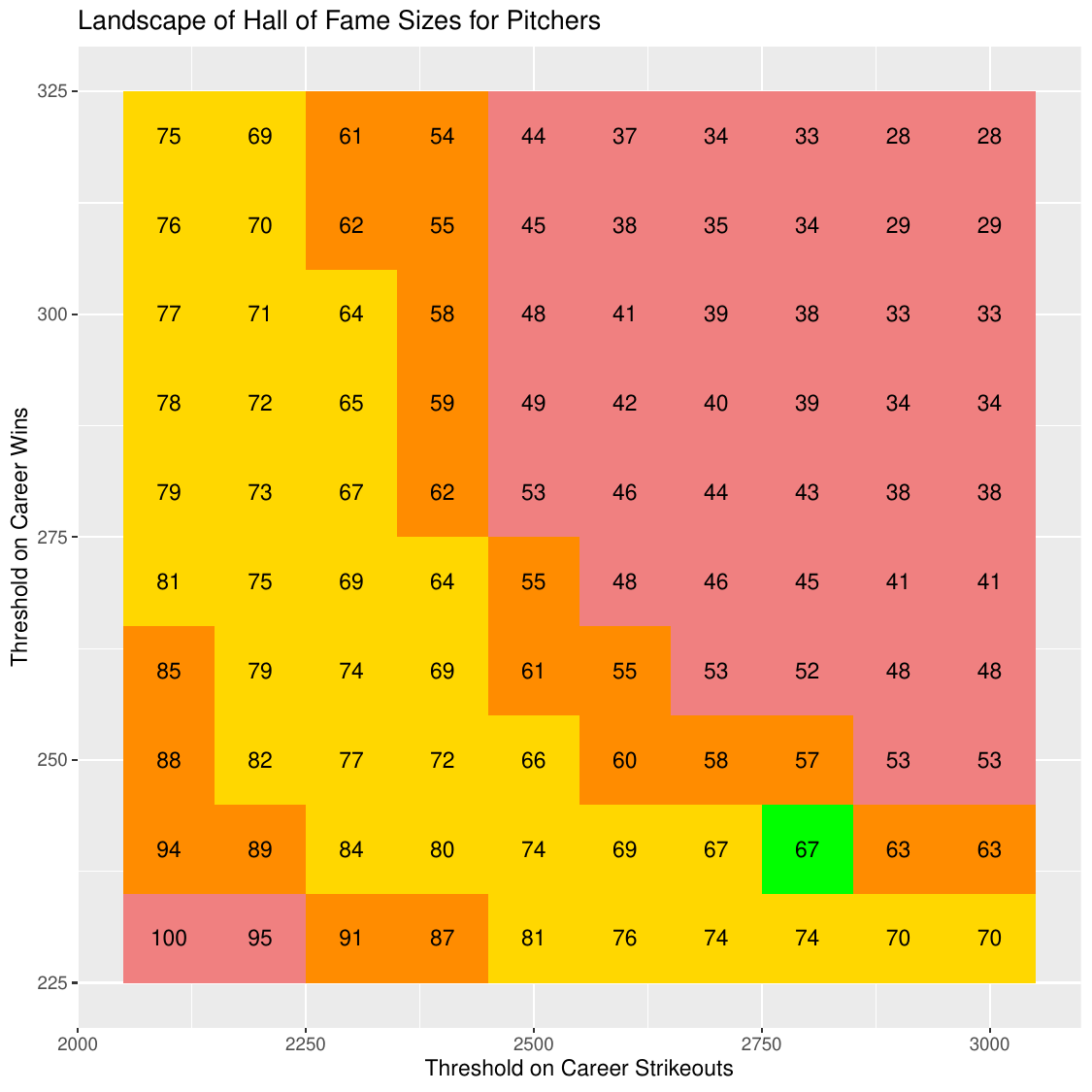}
\centering
\caption{Joint landscape of Hall of Fame sizes for combinations of different thresholds on career strikeouts and wins. Yellow combinations are near to the size the size of the elected HOF (74 pitchers).}
\label{fig-landscape-pitchers}    
\end{figure}

Figure~\ref{fig-landscape-pitchers} gives the joint landscape of HOF sizes for a grid of reasonable thresholds on career strikeouts and wins for pitchers.  The set of possible thresholds was again chosen to span a wide range of elite career values while only considering round numbers so that the resulting Hall of Fame cut is easy to remember. 

We see that many different combinations of thresholds on career strikeouts and wins lead to Hall of Fame sizes that are close to the 74 pitchers in the elected Hall of Fame.  Within those ranges, I chose the thresholds of 2800 strikeouts and 240 wins which leads to a Hall of Fame with 67 major league pitchers. 

I prefer thresholds that give a smaller set of Hall of Fame pitchers (67) compared to the 71 pitchers in the elected Hall of Fame in order to leave some space for relief pitchers that do not have into the Hall of Fame via career strikeouts and wins. I will discuss relief pitchers in more detail in Section~\ref{comparing-pitchers}.   

If we were to raise the threshold on career strikeouts from 2800 to 3000, we would reduce the HOF size from 67 pitchers down to 63 with one elected HOFer, Jim Bunning, among the 4 additional pitchers that would be excluded.  If we were to raise the threshold on career wins from 240 to 250, we would reduce the HOF size substantially from 67 pitchers down to 57 pitchers, with five elected HOFers including Juan Marichal among the 10 additional pitchers that would be excluded.

If we were to lower the threshold on career strikeouts from 2800 to 2600, we would increase the HOF size from 67 up to 69 pitchers, with David Cone and Chuck Finley (neither elected HOFers) being the two pitchers added.  If we were to lower the threshold on career wins from 240 to 220, we would increase the HOF size substantially from 67 pitchers up to 86, with four elected HOFers (Mordecai Brown, Waite Hoyt, Whitey Ford, Catfish Hunter) among the 19 additional pitchers that would be included.

\section{Comparing Batters between HOF Cut and Elected HOF} \label{comparing-batters}

Now that I've established my Hall of Fame Cut, I will compare the set of 165 major league batters that make my Hall of Fame Cut to the 171 major league batters in the elected Hall of Fame.  Just over half, 95 out of 171, of the major league batters in the elected Hall of Fame also make the Hall of Fame Cut given in Figure~\ref{fig-cut}.   

There are 70 major league batters that make the Hall of Fame Cut but that are not in the elected Hall of Fame.  In Table~\ref{top-nonHOF-batters}, I list the top ten batters ranked by career hits and the top ten batters ranked by career home runs that are not currently in the elected Hall of Fame.  

\begin{table}[h]
\begin{center}
\caption{Top ten batters ranked by career hits and the top ten batters ranked by career home runs that are not currently in the elected Hall of Fame.} \label{top-nonHOF-batters}
\begin{tabular}{|lc|lc|}
\hline
Name  & H & Name & HR \\ 
\hline
Pete      Rose & 4256   &   Barry     Bonds  & 762 \\
Albert    Pujols$^*$  &  3384   &  Albert    Pujols$^*$  & 703 \\
Alex Rodriguez  & 3115    &   Alex Rodriguez  & 696 \\
Ichiro    Suzuki$^*$ &  3089   &   Sammy      Sosa  & 609 \\
Miguel   Cabrera$^*$ &  3088  &     Mark   McGwire  & 583 \\
Rafael  Palmeiro &  3020  &   Rafael  Palmeiro  & 569 \\
Barry     Bonds &  2935  &    Manny   Ramirez  & 555 \\
Omar   Vizquel &  2877   &    Gary Sheffield  & 509 \\
Johnny     Damon &  2769  &   Miguel   Cabrera$^*$  & 507 \\
Vada    Pinson  & 2757  &   Carlos   Delgado  & 473 \\
\hline
\end{tabular}
\end{center}
\end{table}

Table~\ref{top-nonHOF-batters} is dominated by players that are not in the elected Hall of Fame because of suspected steroid use, including the all time leader in career home runs Barry Bonds.   The all time leader in career hits, Pete Rose, is also not in the elected Hall of Fame because of a lifetime ban for sports betting.   The transparency of my Hall of Fame cut means it is agnostic to performance enhancement or off field issues.  

There are also several players, indicated by a $^*$ in Table~\ref{top-nonHOF-batters}, that not yet eligible for the Hall of Fame but will certainly get elected: Albert Pujols, Ichiro Suzuki and Miguel Cabrera.  There are also less prominent players that did not come close to election to the Hall of Fame but nonetheless clearly make my Hall of Fame cut.  

One such player is Johnny Damon who makes my Hall of Fame Cut based on his 2769 career hits.  In addition to those hits, Johnny Damon had 235 home runs, 408 stolen bases and a career bWAR of 56.3, yet he did not receive the minimum 5\% of votes to stay on the ballot in his first year of eligibility for the Hall of Fame.    





\begin{figure}[ht!]
\includegraphics[width=5.5in]{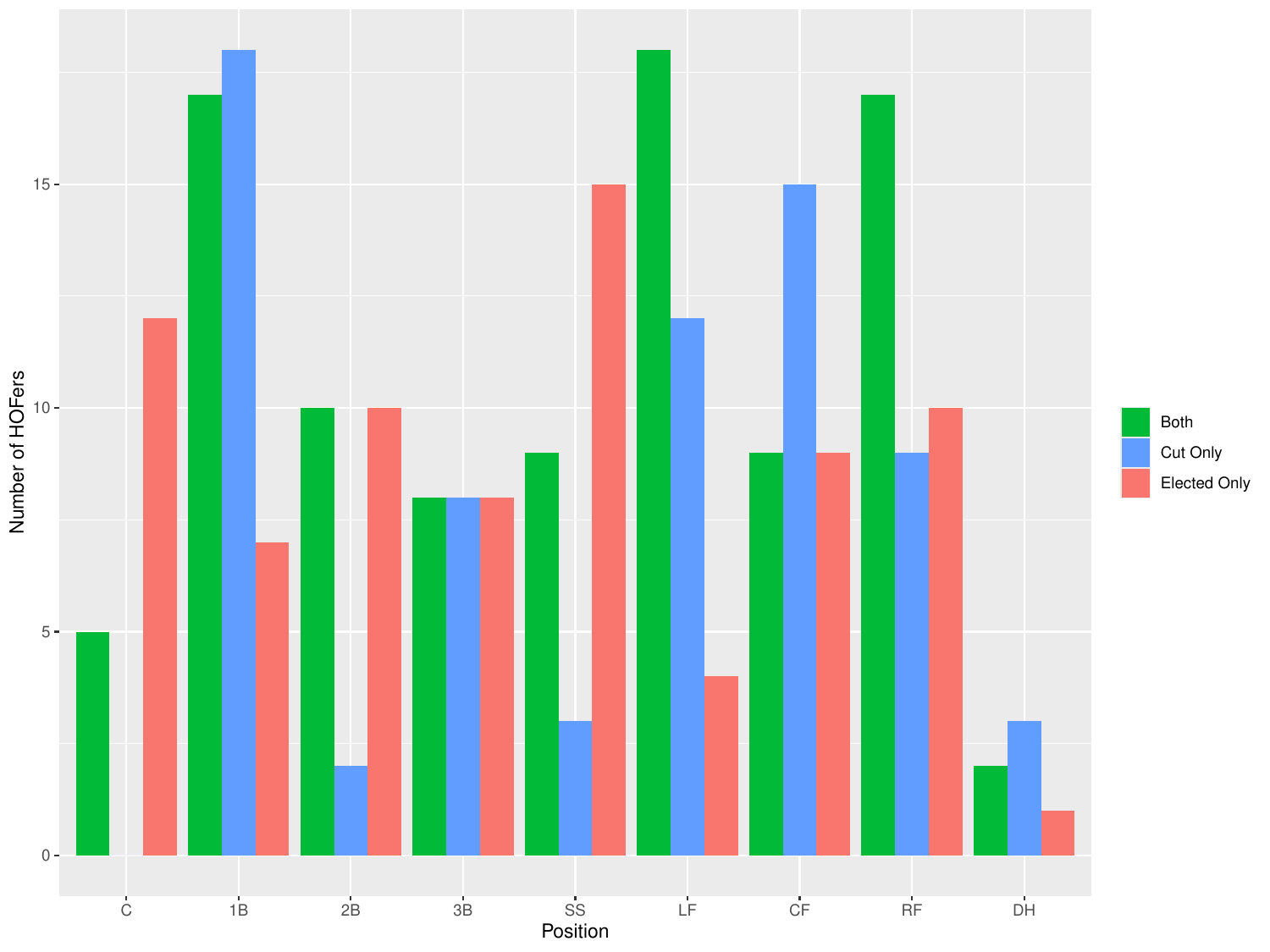}
\centering
\caption{Numbers of batters in 3 different categories grouped by fielding position. Categories are Both = makes HOF Cut and in elected HOF, Cut Only = makes HOF Cut but not in elected HOF, Elected Only = doesn't make HOF Cut but in elected HOF}
\label{fig-barplot-position-batters}    
\end{figure}

Figure~\ref{fig-barplot-position-batters} compares my Hall of Fame Cut versus the elected Hall of Fame for all batters grouped by their fielding position.  My Hall of Fame cut is based on cumulative performance compared to all other batters and so is completely agnostic to fielding position.  However, it is clear from Figure~\ref{fig-barplot-position-batters} that there is substantial positional adjustment applied by voters for the elected Hall of Fame.  

There are fewer catchers, second basemen and shortstops that make the Hall of Fame Cut than have been elected to the Hall of Fame which suggests that voters are implicitly accepting lower batting performance at these important defensive positions.  Interestingly, there are a lot of centerfielders that make the Hall of Fame Cut despite the reputation of CF as a key defensive position.  

Only five out of the 17 catchers in the elected Hall of Fame also make the Hall of Fame cut: Johnny Bench, Yogi Berra, Carlton Fisk, Mike Piazza and Ivan Rodriguez.  Ozzie Smith and Bill Mazeroski are elected Hall of Famers noted for their defense that do not make the Hall of Fame Cut.

\section{Comparing Pitchers between HOF Cut and Elected HOF} \label{comparing-pitchers}

A slightly higher proportion, 46 out of 74, of the major league pitchers in the elected Hall of Fame also make the Hall of Fame Cut given in Figure~\ref{fig-cut}.   There are 21 major league pitchers that make the Hall of Fame Cut but that are not in the elected Hall of Fame.  In Table~\ref{top-nonHOF-pitchers}, I list the top ten pitchers ranked by career strikeouts and the top ten pitchers ranked by career wins that are not currently in the elected Hall of Fame.  

\begin{table}[h]
\begin{center}
\caption{Top ten pitchers ranked by career strikeouts and the top ten pitchers ranked by career wins that are not currently in the elected Hall of Fame.} \label{top-nonHOF-pitchers}
\begin{tabular}{|lc|lc|}
\hline
Name  & K & Name & W \\ 
\hline
Roger   Clemens &  4672   &  Roger   Clemens  & 354 \\
Justin Verlander &  3198  &   Bobby   Mathews  & 297 \\
Max  Scherzer  & 3193   &  Tommy      John  & 288 \\
Curt Schilling &  3116   &   Tony   Mullane  & 284 \\
CC  Sabathia  & 3093   &  Jamie     Moyer  & 269 \\
Zack   Greinke  & 2882    &   Jim McCormick  & 265 \\
Mickey    Lolich  & 2832   &    Gus   Weyhing  & 264 \\
Clayton   Kershaw  & 2807   &   Andy  Pettitte  & 256 \\
Frank    Tanana  & 2773    &    Al  Spalding  & 251 \\
Bartolo     Colon  & 2535   &     CC  Sabathia  & 251 \\
\hline
\end{tabular}
\end{center}
\end{table}

Table~\ref{top-nonHOF-pitchers} is dominated by pitchers that are not yet eligible for the Hall of Fame but will certainly be elected on their first ballot: Justin Verlander, Max Scherzer, CC Sabathia and Clayton Kershaw.  Andy Pettitte and Zach Greinke are not as sure fire but I believe they wlll eventually get elected to the Hall of Fame as well.  We again see that the Hall of Fame Cut is agnostic to suspected steriod use (Roger Clemens) and off field issues (Curt Schilling).   

Just as with Johnny Damon on the batting side, the Hall of Fame Cut also highlights some underappreciated pitchers closer to the margins of the Hall of Fame.  One such pitcher is Bartolo Colon who makes the Hall of Fame Cut based on his 247 wins.  In addition to those wins, Bartolo Colon had 2535 strikeouts and a career bWAR of 46.2, yet he also did not receive the minimum 5\% of votes to stay on the ballot in his first year of eligibility for the Hall of Fame.    



My Hall of Fame cut is based on cumulative performance which has important consequences for pitchers.  The first is there are several pitchers, such as Sandy Koufax and Roy Halladay that were elected to the Hall of Fame for their peak peformance but did not have the longevity to make the Hall of Fame cut.  We will revisit this longevity issue in Section~\ref{toptier} below. 

The other consequence is that relief pitchers have no path to making the Hall of Fame Cut due to their restricted usage.  Even Mariano Rivera, the greatest relief pitcher of all time, only accumulated 1173 strikeouts and 82 wins in his pitching career.  John Smoltz spent more of his career as a starting pitcher and accumulated 3084 strikeouts to make the Hall of Fame Cut but Dennis Eckersley spent a little more of his career as a relief pitcher and misses the Hall of Fame Cut with only 2401 strikeouts. 

My chosen thresholds in Figure~\ref{fig-cut} lead to a smaller set of pitchers compared to the elected Hall of Fame (67 vs. 71) which leaves room to add several relief pitchers if desired.   We could extend my Hall of Fame cut framework to include a third pitching count category specific to relief pitchers.  The most natural choice would be career saves and with a threshold of 500 on career saves, Mariano Rivera and Trevor Hoffman would make this extended Hall of Fame cut.  

\section{The Top Tier of the Hall of Fame} \label{toptier}

I dedicate this next exercise to my colleague Eric Bradlow in honor of his interest in defining an inner circle within the Hall of Fame.  It is easy to repeat my Hall of Fame Cut approach except that I now focus on the landscape of thresholds that limit the set of included players to around 18 batters and 10 pitchers. These numbers are chosen to roughly corresponds to the top 10 \% of players in the Hall of Fame.  

There are several good combinations of thresholds on hits and home runs that define a top tier of around 18 batters.  Among these, my preferred top tier thresholds are H $> 3300$ or HR $> 600$ which defines a set of 19 batters in the ``Top Tier HOF Cut".  The full list of these inner circle batters and their hit and home run counts are given in Table~\ref{top-tier-table}.

The threshold combination of H $> 3500$ or HR $> 550$ also defines a set of 19 batters but this alternative top tier gives even more preference to steroid-era power batters.  Specifically, Mark McGwire, Manny Ramirez, Rafael Palmeiro and Reggie Jackson make this alternative top tier while Cap Anson, Carl Yastremski, Paul Molitor and Derek Jeter do not make this alternative top tier. 

There are also several good combinations of thresholds on strikeouts and wins that define a top tier of around 10 pitchers.  Among these, my preferred top tier thresholds are K $> 4000$ and W $> 350$ which defines a set of 12 pitchers in the ``Top Tier HOF Cut".  The full list of these inner circle pitchers and their strikeout and win counts are also given in Table~\ref{top-tier-table}.

We see an even split between six early era pitchers and six top tier modern pitchers: Warren Spahn, Steve Carlton, Nolan Ryan, Roger Clemens, Greg Maddux and Randy Johnson.  It is difficult to argue against any of these pitchers as inner circle Hall of Famers.  However, it is also difficult to consider any list of inner circle pitchers that does not include Sandy Koufax and Pedro Martinez, the two players with the greatest peak pitching performance in baseball history. 

Neither Sandy Koufax nor Pedro Martinez had the longevity necessary to make a top tier threshold based on career counts, with Koufax not even making the Hall of Fame Cut in Figure~\ref{fig-cut}.  It is an inherent limitation of our count-based focus, though any consideration of a rate-based statistic would have its own issues as I discuss in Section~\ref{whatstats} above.  

The most glaring omission on the batting side is Ted Williams whose rate-statistic performance (all time leader with .482 OBP, min 3000 PA) would certainly qualify as inner circle Hall of Fame but whose counts were reduced by missing almost five full seasons serving in World War II and Korea.  

\bigskip

\begin{table}[h]
\begin{center}
\caption{The 19 batters and 12 pitchers that make the Top Tier HOF Cut: H $> 3300$ or HR $> 600$ or K $> 4000$ or W $> 350$. Batters and pitchers are separately listed in order of career bWAR \citep{war}.} \label{top-tier-table}
\begin{tabular}{|lrrr|lrrr|}
\hline
Name  & H  & HR & bWAR & Name  & K  & W & bWAR \\ 
\hline
Babe Ruth & 2873 & 714 & 182.6 & Walter Johnson & 3509 & 417 & 166.9 \\ 
Barry Bonds & 2935 & 762 & 162.8 & Cy Young & 2803 & 511 & 163.6 \\
Willie Mays & 3293 & 660 & 156.2 & Roger Clemens  & 4672 & 354 & 139.2\\
Ty Cobb   & 4189 & 117 & 151.5 & Grover Alexander & 2198 & 373 & 119.6 \\
Hank Aaron & 3771 & 755 & 143.1 & Kid Nichols & 1881 & 362 & 116.3  \\ 
Tris Speaker & 3514 & 117 & 134.9 & Christy Mathewson & 2507 & 373 & 106.7 \\
Honus Wagner & 3420 & 101 & 131.0 & Greg Maddux  & 3371 & 355 & 106.6\\
Stan Musial  & 3630 & 475 & 128.5 & Randy Johnson  & 4875 & 303 & 101.1\\
Eddie Collins  & 3315 & 47 & 124.3 & Warren Spahn  & 2583 & 363 & 100.1\\
Alex Rodriguez & 3115 & 696 & 117.6 & Steve Carlton & 4136  & 329 & 90.2 \\
Albert Pujols & 3384 & 703 & 101.4 & Nolan Ryan & 5714 & 324 & 81.3 \\
Carl Yastrzemski & 3419 & 452 & 96.5 & Pud Galvin & 1807 & 365 & 73.4  \\
Cap Anson & 3435 & 97 & 94.3 & & & & \\
Ken Griffey Jr. & 2781 & 630 & 83.8 & & & & \\
Pete Rose & 4256 & 160 & 79.5 & & & & \\
Paul Molitor & 3319 & 234 & 75.6 & & & & \\
Jim Thome & 2328 & 612 & 73.1 & & & & \\
Derek Jeter & 3465 & 260 & 71.3 & & & & \\
Sammy Sosa& 2408 & 609  & 58.6 & & & & \\
\hline
\end{tabular}
\end{center}
\end{table}

\section{Hall of Fame Cut for Recent and Active Batters}  \label{recent-and-active-batters}

As noted above, there are several recently retired batters that make my Hall of Fame Cut but have not {\it yet} been elected to the Hall of Fame. 
Albert Pujols not only makes the Hall of Fame Cut but also makes the top tier Hall of Fame Cut in Section~\ref{toptier} with 3384 hits and 703 home runs.
Miguel Cabrera makes the Hall of Fame Cut with 3174 hits and 511 home runs. 

Ichiro Suzuki makes the Hall of Fame Cut with 3089 hits which is especially notable given that he played several years of professional baseball in Japan before starting his major league career.  Pujols, Cabrera and Suzuki are certain to be elected to the Hall of Fame once they are eligible (5 years post-retirement).

Carlos Beltran also makes the Hall of Fame Cut with 2725 hits and he is probable but not certain to be elected to the Hall of Fame.  Robinson Cano makes the Hall of Fame Cut with 2725 hits but was banned for a season because of steroid usage and unlikely to be elected to the Hall of Fame. Nelson Cruz (464 home runs) and Joey Votto (356 home runs) both make the Hall of Fame Cut but neither are likely to be elected to the Hall of Fame.

Another positive aspect of my Hall of Fame Cut is that players achieve the standard (and can be celebrated for it!) while they are still active players.  
At the end of the 2024 season, three active batters have career home runs above the Hall of Fame Cut:  Giancarlo Stanton (429 HR), Mike Trout (378 HR) and Paul Goldschmidt (362 HR).   

There are no active batters that make the HOF Cut based on career hits but Freddie Freeman (2267 hits) and Jose Altuve (2232 hits) are likely to achieve the standard within the next few years.  Several batters are also likely to make the Hall of Fame Cut based on their current home run totals: Freddie Freeman again (343 HR), Manny Machado (342 HR), Nolan Arenado (341 HR), Bryce Harper (336 HR) and Aaron Judge (315 HR).  

\section{Hall of Fame Cut for Recent and Active Pitchers}  \label{recent-and-active-pitchers}

There are also several recently retired pitchers that make my Hall of Fame Cut but have not {\it yet} been elected to the Hall of Fame. CC Sabathia is certain to be elected to the Hall of Fame once eligible, and he makes the Hall of Fame Cut based on either his 251 wins or his 3093 strikeouts.  Zack Greinke makes the Hall of Fame Cut based on his 2979 strikeouts and is less certain but likely to be elected to the Hall of Fame.  

I already highlighted Bartolo Colon as making my Hall of Fame Cut with 247 wins but not based on his 2535 strikeouts.  Recently retired pitcher Felix Hernandez has almost the same number of career strikeouts (2524 K) but played fewer seasons on worse teams and finished his career with only 169 wins, well below the Hall of Fame Cut.

At the end of the 2024 season, three active pitchers have career strikeouts above my Hall of Fame Cut: Justin Verlander (3416 K), Max Scherzer (3407 K), and Clayton Kershaw (2968 K).  Justin Verlander is the only active pitcher with career wins (262 W) above my Hall of Fame Cut.  

If current trends in pitching usage continue, it is unlikely that any other pitcher will achieve the Hall of Fame cut of 240 wins, which is already much lower than the more traditional benchmark of 300 wins for the Hall of Fame.  Randy Johnson will probably be the last pitcher ever to retire with more than 300 wins and Justin Verlander will probably be the last pitcher ever to retire with more than 250 wins.  

Gerrit Cole is currently the next highest active pitcher with 153 wins but is unlikely to get to 240.  However, Cole currently has 2251 strikeouts and is likely to make the threshold of 2800 with a few more healthy seasons.   Chris Sale is even closer with 2414 career strikeouts but less likely than Cole to have several more healthy seasons.  Longer term but worth tracking is Aaron Nola who already has 1779 career strikeouts at the age of 31. 

\section{Discussion} \label{discussion}

My Hall of Fame Cut provides a ``bar napkin simple" standard for career greatness in baseball: H $> 2500$ or HR $> 350$ or K $> 2800$ or W $> 240$.   
Similar to the Golden Players Club for Japanese baseball, I focus on counting measures of career performance but I use four career totals (two each for batting and pitching) to capture career greatness in major league baseball.  

More complex system such as Jay Jaffe's JAWS \citep{jaws} can provide a more nuanced summary of a player's career but lacks the transparency and simplicity that needed for a quantitative historical standard.  Rate statistics (such as OBP or ERA) can be better indicators of greatness on the seasonal level but they can not be used here without setting additional thresholds to deal with small sample players.  Additionally, we want a standard that is permanent once achieved which would be difficult with a rate statistic.  

I choose career hits and home runs to represent the two extremes of the batting power spectrum which provides a path to the Hall of Fame for both great contact and power hitters.  I chose career strikeouts as the most fundamental unit of isolated pitching performance.  I chose wins in deference to their use as historical benchmarks of pitching greatness.  I have not included playoff performance in my analysis in deference to historical norms (e.g. playoff wins are not counted when discussing a pitcher as won 300 games or not) but one could use my same approach to construct a HOF cut with playoff counts included.

Seperately for batters and pitchers, I explored the landscape of Hall of Fame sizes resulting from different thresholds on those four career counts and chose thresholds to give roughly the same number of major league players (67 pitchers and 165 batters) in my Hall of Fame cut compared to the elected Hall of Fame (74 pitchers and 171 batters).   This same approach was also used to define a more stringent ``top tier" Hall of Fame Cut (H $> 3300$ or HR $> 600$ or K $> 4000$ or W $> 350$) that represents the top 10\% of historical greatness for pitchers and batters. 

The majority of major league batters (95 out of 171) and pitchers (46 out of 74) in the elected Hall of Fame also make my Hall of Fame Cut. 
However, there are an additional 70 batters and 21 pitchers that make the Hall of Fame Cut but are not yet in the elected Hall of Fame.  This provides an opportunity to highlight some under-appreciated players such as Johnny Damon and Bartolo Colon.  

A benefit of my Hall of Fame Cut is that we can celebrate players right as they achieve the standard rather than having to wait on a lengthy HOF voting process.  We can celebrate the recently retired Albert Pujols, Miguel Cabrera, Ichiro Suzuki and CC Sabathia as Hall of Fames even though they are not yet eligible for HOF voting.  We can also celebrate the following active (as of 2024) players as already having made the Hall of Fame Cut: Giancarlo Stanton, Mike Trout, Paul Goldschmidt, Justin Verlander, Max Scherzer, and Clayton Kershaw. 

The Hall of Fame Cut is agnostic to the performance enhancement or off field issues that have kept several of the greatest batters and pitchers of all time out of the elected Hall of Fame.  The same cumulative standard is applied to all batters which is a stark contrast to the elected Hall of Fame where fielding position is a major focus of voters.   The catching position is especially disadvantaged with only 5 catchers making the Hall of Fame Cut compared to 17 catchers in the elected Hall of Fame.  

The same cumulative standards for all pitchers means that relief pitchers have no path to making the Hall of Fame Cut due to their restricted usage.  Mariano Rivera is the greatest of all time in that role but pitched less than 1300 innings in his career, whereas the average Hall of Fame starting pitcher has pitched almost three times as many innings ($\approx$3500).   One could correct this issue by adding saves as a third pitching count to the Hall of Fame Cut, in which case I personally favor a relatively high threshold of 500 saves which has so far only been achieved by Mariano Rivera and Trevor Hoffman.  

Wins were included in my Hall of Fame cut in deference to their history as a benchmark but they are becoming less relevant over time as a measure of pitching greatness, and it is unlikely that any more pitchers will come close to the threshold of 240 wins.   So there is a need for additional measures of pitching performance (such as quality starts) even if they have the disadvantage of not having been recorded for all of baseball history.  

The Hall of Fame Cut also naturally favors consistency and longevity over peak performance.  Sandy Koufax had one of the greatest peaks of all time for starting pitchers but did not have the longevity to accumulate enough wins or strikeouts to make the Hall of Fame Cut.   Additional measures could also be considered if we wanted to capture other isolated aspects of baseball performance such as base stealing.  However, the goal of this analysis was not to reconstruct all the complexities of baseball performance or history, but rather to see how well we could capture baseball greatness with a limited set of rules (Figure~\ref{fig-cut}) that you've probably memorized by this point.  

\bibliographystyle{natbib}

\bibliography{references}

\end{document}